\documentclass[11pt]{article}
\usepackage{epsfig}
\usepackage{times}

\usepackage{epsfig}
\begin{document}                
\newcommand{\keywords}{02.40.Dr Euclidean and projective geometries;
03.75.-b Matter waves; 04.30.Db Wave generation and sources;
11.10.Cd Axiomatic approach.}
\pagestyle{myheadings} \markright{Prospects for unification under 4DO}
\title{Prospects for unification under 4-dimensional optics}
\author{Jos\'e B. Almeida\footnote{Universidade do Minho,
Departamento de F\'isica, 4710-057 Braga, Portugal. E-mail:
\texttt{bda@fisica.uminho.pt}}}

\date{}
\maketitle
\begin{abstract}                
4-dimensional optics is here introduced axiomatically as the
space that supports a Universal wave equation which is applied to
the postulated Higgs field. Self-guiding of this field is shown
to produce all the modes necessary to provide explanations for
the known elementary particles. Forces are shown to appear as
evanescent fields due to waveguiding of the Higgs field, which
provide coupling between waveguides corresponding to different
particles. Carrier particles are also discussed and shown to
correspond to waveguided modes existing in 3-dimensional space.

\noindent{\hrulefill}

\noindent{\textbf{PACS: }\keywords}

\end{abstract}
\section{Introduction}
Relativity and quantum mechanics seem to be inconsistent theories
nonetheless because the geometries used are entirely different.
While relativity is set in Riemannian geometry, with signature
$(-,+,+,+)$, quantum mechanics employs Euclidean geometry with
signature $(+,+,+,+)$. Several approaches have been proposed to
circumvent the problem; Kitada \cite{Kitada94, Kitada96} proposed
an orthogonalized space of dimension 10, obtained by
multiplication of the relativistic Riemannian space by the
Euclidean phase space of quantum mechanics, complemented with the
local time concept. It is well known, though, that superstring
theory provides today the most promising prospects for
unification, making use of 11-dimension space \cite{greene99}.

The author has shown that relativistic problems can conveniently
be expressed in Euclidean space where proper time, rather than
time, is used as 0th coordinate \cite{Almeida01:4, Almeida02}. He
has also shown that inertial mass can be associated to a guided
wave on that space, whereas gravity appears as an evanescent
field due to that same guided wave \cite{Almeida01:5}; this
theory has been baptized as 4-dimensional optics, henceforth
designated by 4DO. Thus there is scope for the construction of a
unified theory which does not suffer from the inconsistency
referred above and 4DO appears as a likely candidate.

The present paper aims to show that not only quantum mechanics
and general relativity but also particle physics, as is presently
explained by the standard model, have a good prospect of being
none but particular aspects of the unified 4DO theory established
on very simple principles. Not all the consequences of the theory
can be derived in this work nor does the author have the
knowledge to draw from the principles he proposes all the
implications. Those that he can, though, are important enough to
give him the courage to sustain that 4DO holds a great potential
for being the next comprehensive theory in physics.

In his previous work the author chose to present 4DO as an
alternative to general relativity, which could also accommodate
quantization. In this paper the starting point is more
fundamental, as the author just assumes that in a given space
there is but one harmonic field submitted to a wave equation whose
solutions are guided modes corresponding to the elementary
particles, these generating evanescent fields which are
responsible for particle interactions.

As stated previously, 4DO space is characterized by the 3 physical
space coordinates, $x$, $y$ and $z$, complemented by coordinate
0, which is clearly defined in Ref.\ \cite{Almeida02} and is
frequently associated with the letter $\tau$; this coordinate is
designated by the name proper time, for reasons that have already
been explained \cite{Almeida01:4, Almeida02}. 4DO space
characterization is completed by the statement that it is an
Euclidean space with signature $(+,+,+,+)$.

This work will make use of three extra dimensions of a localized
nature. The local characteristic of these extra dimensions
justifies that the theory maintains its initial name of
4-dimensional optics, for most of what is perceived by observers
can be set in 4-dimensional space. The new dimensions are
sometimes associated with a rigid body's Euler angles and will be
designated by the greek letters $\iota$, $\kappa$ and $\lambda$.

4DO uses non-dimensional units, due to the frequent appearance of
coordinates in exponents and the inconvenience of constants being
present in most equations. Non-dimensional units are obtained
dividing length, time and mass by the normalizing factors $\sqrt{G
\hbar /c^3}$, $\sqrt{G \hbar/c^5}$ and $\sqrt{\hbar c/G}$,
respectively, while electric charge is normalized by the proton's
charge. The speed of light in vacuum is $c$, $G$ is the
gravitational constant and $\hbar$ is Planck's constant divided
by $2 \pi$. In non-dimensional units all entities are expressed
by pure numbers, which in itself raises interesting philosophical
questions worthy of discussion in a different sort of paper.
\section{Founding principles}
The unified 4DO theory is based on the assumption that all
elementary particles are manifestations of self-guided modes of a
Universal field called the Higgs field with a Universal frequency
designated by Higgs frequency. For a stationary particle it is
believed that the Higgs field generates a standing wave pattern
laid along the $\tau$ axis and more generally, for non-stationary
particles, the standing wave pattern is laid along the particle's
worldline. The basis for the previous assumption was established
in Ref.\ \cite{Almeida01:5} for the derivation of mass and
gravity and is here extended to the generation of elementary
particles themselves.

In a similar way to which gravity results from the evanescent
field of guided waves \cite{Almeida01:5}, it is believed that all
the fundamental interactions in nature must be the observable
effects of evanescent fields. Whether self-guided or guided by
external fields, all guided waves generate evanescent fields
which are believed to explain all the four fundamental
interactions.

In the above cited work the author associated an elementary
particle's field $\Psi$ to the Lagrangian density
\begin{equation}
    \label{eq:lagdens}
    \mathcal{L} = \frac{n^{\alpha \beta}
    \partial_\alpha \Psi \partial_\beta \Psi}{\Psi \Psi^*} -
    \Psi^2,
\end{equation}
from which one could derive the Euler-Lagrange equation
\begin{equation}
    \label{eq:wave}
    \partial_\alpha \left(\frac{n^{\alpha \beta}}{ \Psi
    \Psi^*}\,\partial_\beta \Psi \right)=
    \partial_\alpha \partial^\alpha \Psi = -\Psi.
\end{equation}
In Cartesian coordinates $n^{\alpha \beta}$ is the metric field
factor, which is complemented by the inertial factor $\Psi
\Psi^*$. It will be shown that when dealing with gravity the
inertial factor becomes the inertial mass, while electric charge
is the inertial factor associated with electric field.

The assumption that the elementary particles result from
self-guided modes of the Higgs field suggests that one should use
an equation based on Eq.\ (\ref{eq:wave}) without the external
field component
\begin{equation}
    \label{eq:fundwave}
    \delta^{\alpha \beta} \partial_\alpha \left(
    \frac{\partial_\beta \Psi }{ \Psi
    \Psi^*}\right) = -\Psi,
\end{equation}
valid in Cartesian coordinates. Nevertheless, the use of Euler
angles as coordinates imposes adaptations to the equation above,
which will be discussed further along. One important
simplification arises from the consideration of stationary
particles, for these have a worldline coincident with the $\tau$
axis. According to the theory developed in Ref.\
\cite{Almeida01:5} the frequency along the worldline equals the
particle's inertial mass $m$, whereby it is legitimate to
postulate a $\tau$ dependence of the type
$\mathrm{e}^{\mathrm{j}m \tau}$, with $\mathrm{j}=\sqrt{-1}$.
\section{Guided modes and elementary particles}
4DO theory is, in many respects, the natural extrapolation of
3-dimensional optical fibers' waveguide theory into 4-dimensions.
Several references will be made to this theory and the reader is
referred to textbooks such as Refs.\ \cite{Okamoto00, Unger77}
for details. Ii will also be useful to associate elementary
particles to rigid bodies, namely for the understanding of
coordinates and degrees of freedom; several textbooks are
available for reference but see for instance Ref.\ \cite{Jose98}.

In optical waveguide theory one usually solves the Helmoltz
equation which is written in Cartesian coordinates as
\begin{equation}
    \label{eq:Helmoltz}
    \delta^{i j} \partial_{i j} \Psi = - k^2 n^2 \Psi;
\end{equation}
here $k$ is the wave number, $n$ is the refractive index and the
indices take values between 1 and 3. Waveguiding arises when $n$
is a function of the coordinates with certain characteristics,
namely when it is a radially decreasing function of some given
axis. Comparing with Eq.\ (\ref{eq:fundwave}), it is clear that
waveguiding can occur without the intervention of a refractive
index because the field amplitude $\Psi \Psi^*$ is inside the
first derivative. In other respects, though, self-guiding is not
very different from waveguiding by means of a refractive index
and is not specific to 4-dimensional space.

The extra dimension in 4DO introduces waveguiding possibilities
that one does not encounter in optical waveguides; the situation
is similar to the rotations of a rigid body compared to those of
a disc. In the latter case the rotation axis is always normal to
the disc and can be made to coincide with the $z$ axis of
cylindrical coordinates, while with a rigid body rotation can
happen about any of the 3 principal axes and is always orthogonal
to the $\tau$ axis of 4-dimensional space for stationary bodies.
Transporting the discussion to the waveguide situation, one is
allowed to make the axis of a cylindrical optical waveguide
coincident with the $z$ axis, so that the angular dependence of
the field can be expressed by the azimuth angle of cylindrical
coordinates. In a 4-dimensional waveguide one makes the waveguide
axis coincide with the $\tau$ axis but needs 3 angles to describe
fully the angular dependence of $\Psi$; these are the Euler
angles $\iota$, $\kappa$ and $\lambda$. In order to use Euler
angles as coordinates one has to consider a 4-dimensional frame
whose 3 spatial axes would coincide with the particle's principal
axes, if the particle were associated to a rigid body.

Not unlike their 3-dimensional counterparts, 4-dimensional
waveguides exhibit modes because the waves must interfere
constructively after a "full rotation", although one must be
careful about the meaning of this expression in 4-dimensional
space. Similar also to optical fibers, 4-dimensional waveguides
can loose and gain modes when they are bent. This process has
already been shown to explain the exchange of gravitons between
orbiting bodies \cite{Almeida01:5} and it will be shown later to
explain decaying of excited atoms through photon emission. In
both these cases one is dealing with wave guidance by fields
rather than self-guidance but there is no reason for a similar
process not to occur with self-guided modes. One has to consider
that an elementary particle is associated with a waveguide whose
axis lays along its worldline, so that all acceleration suffered
by the particle is translated into bending of the associated
waveguide and consequent mode loss. This justifies that only
first order modes can be considered stable because higher order
modes will decay naturally into the lower order ones.

This paragraph looks at the possible first order mode solutions
of Eq.\ (\ref{eq:fundwave}) in order to show that indeed they
match qualitatively all the elementary particles that make the
standard model \cite{Barnett00, Halzen84}.
\begin{description}
\item [Electron, muon and tau:]
The electron corresponds to a spherically symmetrical mode
obtained by simultaneous rotation about the 3 principal axes. One
can consider the variable $\sigma = 2(\iota + \kappa +
\lambda)/3$ as an independent variable with period $2\pi$ and
write
\begin{equation}
    \label{eq:electron}
    \Psi = f(r) \mathrm{e}^{\mathrm{j} (\pm  \sigma +  m \tau )}.
\end{equation}
The equation establishes that the guided mode must have a proper
time frequency equal to the electron's mass, here designated by
$m$. After one full rotation of the angular variable $\sigma$ the
function $\Psi$ is replicated, ensuring that this is the lowest
order mode. The plus and minus signs on the exponent denote
rotation in the two possible directions and can be assigned to the
electron and positron, respectively. The electric charge is
associated with the mode number and thus, if there are higher
order modes, even short-lived ones, they must be associated with
particles having an electric charge multiple of the electron's.

The heavier members of the family, the muon and the tau, must be
described also as first order modes; it is believed that $f(r)$
holds the key to understanding the different families of
particles. It is likely that there are different solutions
corresponding to the same Higgs wave orbiting around a core at
different distances from the center, the larger the distance the
bigger the particle's mass.
\item [The quarks:]
In some respects quarks are simpler particles than electron and
its family members because they are associated with rotations
about one or two axes. Starting with the down quark, it is
associated with a rotation about a single axis and is described by
the equation
\begin{equation}
    \label{eq:down}
    \Psi = f(r) \mathrm{e}^{\mathrm{j} (\pm  \iota +  m \tau )},
\end{equation}
where it has been assumed that rotation was about the axis
associated with Euler angle $\iota$ and $m$ is the down quark's
mass.

This equation is very similar to the electron's
(\ref{eq:electron}) but hides some important differences. Notice
that the total charge of the down quark can be seen as $1/3$ the
electron's isotropic charge but with a preferred orientation
along one of the principal axes. This preferred orientation is
assigned to the quark's color and the three axes are naturally
associated with the three quark colors. As with the electron, one
of the rotation directions is related to the quark and the other
to the anti-quark.

The up quark is similar to the down quark in every respect,
except that simultaneous rotation around two axes must be
considered. Naturally the electric charge has now an absolute
value of $2/3$, with both signs possible. In this case the plus
sign is associated with the quark and the minus sign with the
anti-quark. Notice that this fact gives color an equivalent
interpretation as for the negative charge in the case of rotation
around one single axis.
\item [The neutrinos:]
The modes associated with the neutrinos don't have correspondents
in optical fibers. They are purely transverse modes which can only
exist in waveguides of infinite longitudinal length or waveguides
that are closed on themselves. The waveguides associated with
elementary particles are either infinite in length or closed,
depending on the Universe being open or closed, and allow the
establishment of these modes. Neutrinos can not be associated
with a worldline in the same sense as massive particles because
for the latter the worldline coincides with the wavefront normal
in the center of the waveguide. As a consequence neutrinos fail
the basic equation of 4DO $(\mathrm{d} t)^2 = n_{\alpha \beta}
\mathrm{d} x^\alpha \mathrm{d} x^\beta$ \cite{Almeida01:4}. The
equation defining the neutrino is quite simply
\begin{equation}
    \label{eq:neutrino}
    \Psi = f(r).
\end{equation}

Just as with electrons, for quarks and neutrinos the different
solutions for $f(r)$ should explain at least three families of
particles.
\end{description}
Solving the differential equations is a  very difficult task and
the author does not have, for the moment, any solutions to
propose for any of the fundamental particles. The neutrino
equation is probably the easiest one to write due to the fact
that the field depends only on $r$. Start by noting that Eq.\
(\ref{eq:fundwave}) can be simplified whenever $\Psi$ is real
\begin{equation}
    \label{eq:wavereal}
    \delta^{\alpha \beta} \partial_{\alpha \beta}
    \left(\frac{1}{\Psi} \right) = -\Psi.
\end{equation}
The first member of the equation is a 4-dimensional Laplacian of
the inverse of $\Psi$. Replacing $\Psi$ by  Eq.\
(\ref{eq:neutrino}) and evaluating the Laplacian in spherical
coordinates one gets the neutrino differential equation
\begin{equation}
    \label{eq:dneutrino}
    \partial_{rr} f +\frac{2 \partial_r f}{r}-
    \frac{2 \left(\partial_r f \right)^2}{f} = -
    f^3,
\end{equation}
which is a non-linear equation unknown to the author.

The electron and quark equations are separable and will eventually
fall into the same radial equation, which will explain why all the
particle families have the same 3 or more levels, corresponding to
different solutions of this equation.

Carrier particles (bosons) are generated in a totally different
manner and their discussion must be postponed until later in this
work.
\section{Evanescent fields and particle interaction}
All guided waves generate evanescent fields outside the
waveguide. This is known for optical waveguides and must also be
true in 4 dimensions, be it for potential generated waveguides or
for self-guided waves. This argument was used in Ref.\
\cite{Almeida01:5} to derive the gravitational field as the
evanescent field associated with the $\tau$ component of the wave
vector. Apart from neutrinos, which don't generate an associated
evanescent field, all particles must generate a gravitational
field, which must be complemented by an evanescent field
associated with isotropic rotation. Quarks will also generate a
field associated with non-isotropic rotation.

The argument for gravitational field derivation is reproduced
here. In Ref.\ \cite{Almeida01:5} It was argued that the radially
symmetric component of the evanescent field $\psi$ due to a
stationary particle of mass $m$ must exhibit a frequency $m$
along the $\tau$ direction so that $\partial_\tau \psi =
\mathrm{j} m \psi$. The field must verify the equation
\begin{equation}
    \label{eq:genwave}
    v^2 \delta^{i j} \partial_{i j} \psi = \partial_{\tau \tau}\psi,
\end{equation}
where the indices take values between 1 and 3 and the letter $v$
designates a generalized propagation speed of wavefronts defined
generally as the derivative $\mathrm{d} s /\mathrm{d} \tau$, with
$\mathrm{d} s$ the arc length of the wavefront normal in flat
Euclidean space.

Naturally the resulting field must have spherical symmetry, which
implies that the wave equation will have a more manageable form
in spherical coordinates. Furthermore, because  $\psi$ is a
function of $r$ and $\tau$ alone, it is possible to express $v$ as
\begin{equation}
    \label{eq:genveloc}
    v = \frac{\partial_\tau \psi}{\partial_r \psi}.
\end{equation}

The operator $\delta^{i j} \partial_{i j}$ is a Laplacian;
considering spherical symmetry one can make the replacement
$\delta^{i j}\partial_{i j} =\partial_{rr} + 2
\partial_r  / r $. Re-writing Eq.\ (\ref{eq:genwave})
in spherical coordinates and inserting Eq.\ (\ref{eq:genveloc})
one gets upon simplification
\begin{equation}
    \label{eq:gravwave}
    \psi
     \left(\partial_{r r} \psi + 2 \frac{\partial_r \psi}{ r} \right)
     = \left(\partial_r \psi \right)^2,
\end{equation}
which has the general solution
\begin{equation}
    \label{eq:gravwave2}
    \psi = C_1 \mathrm{e}^{(C_2/r \pm \mathrm{j}\, m \tau)}.
\end{equation}

So far  no comments were made about the nature of the field $\psi$
but this question must be addressed in order to understand its
relation to gravity. It is postulated that $\psi$ is the local
coordinate scale factor, by which it is  meant that space is
corrugated with the Compton frequency on the particle's worldline
and that this corrugation is extended to infinity on the form of
an evanescent field. There must be a transition from the field on
the worldline to the evanescent field but so far there are no
means to choose among the many possibilities. In any case a
particle will always act as a 4-dimensional waveguide for the
field $\psi$, which will allow the extrapolation of many effects
known in their 3-dimensional counterparts.

The field $\psi$ defines the local scale factor or alternatively
it defines how the geodesic arc length should be measured;
accordingly one makes the assignment $g_{\alpha \beta}=\psi
\psi^* \delta_{\alpha \beta}$ for the metric around the particle.
In the absence of mass it is expected that the scale factor will
be unity and so constant $C_1$ in the equation above can be made
unity; constant $C_2$ must become zero for zero mass. The actual
value for constant $C_2$ is easy to establish resorting to
compatibility with Newton mechanics; if this path is taken
constant $C_2$ can be made equal to the mass, in a similar way to
what was used in Refs.\ \cite{Almeida01:5, Almeida01:2}. This
argument will be used in the present work and an independent
derivation of this constant's value will be deferred until there
is better understanding of the waveguiding process. Consequently
the gravitational field due to a stationary elementary particle
will be written as
\begin{equation}
    \label{eq:elementgrav}
    \psi = \mathrm{e}^{(m/r \pm \mathrm{j}\,m \tau)}.
\end{equation}

A similar argument holds for the field associated with isotropic
charge, except that the field is now a function of $r$ and
$\sigma$. One has to replace $\partial_\tau \psi$ by
$\partial_\sigma \psi$ and assume a $\sigma$ dependence such that
$\partial_\sigma \psi = \mathrm{j} q \psi$, where $q$ represents
the particle's electric charge. The end result is obviously an
evanescent field given by
\begin{equation}
    \label{eq:elementcharge}
    \psi = \mathrm{e}^{(q/r \pm \mathrm{j}\,q \sigma)}.
\end{equation}

Although the fields resulting from Eqs.\ (\ref{eq:elementgrav})
and (\ref{eq:elementcharge}) have similar expressions, they have
quite different effects on the space around the particle, due to
the fact that gravity depends on $\tau$ and the electric field
depends on $\sigma$. This aspect will be discussed further along.

One has to consider finally the evanescent field originated by
non-isotropic rotation, for instance that associated to a down
quark. The rotation takes place around an axis which one can
consider aligned with the $x$ axis for the sake of the argument.
Under these circumstances the $x$ axis becomes an axis of
symmetry and on this axis the field must depend only on one of
the Euler angles, $\iota$ for instance, and the value of the $x$
coordinate. Re-writing Eq.\ (\ref{eq:genwave}) with the
appropriate modifications
\begin{equation}
    \label{eq:iotawave}
    v^2 \delta^{i j} \partial_{i j} \psi = \partial_{\iota \iota}\psi,
\end{equation}
with $\partial_\iota \psi = \mathrm{j}\psi$.

On the $x$ axis, for the reasons stated above, $v$ can be replaced
by
\begin{equation}
    \label{eq:xveloc}
    v = \frac{\partial_\iota \psi}{\partial_x \psi},
\end{equation}
which can be introduced in Eq.\ (\ref{eq:iotawave}). In Cartesian
coordinates it is
\begin{equation}
    \label{eq:iotawave2}
    \psi \partial_{x x} \psi = \left(\partial_x \psi \right)^2.
\end{equation}
Solving the equation one gets the particular solution
\begin{equation}
    \label{eq:elementcolor}
    \psi = \mathrm{e}^{j(x \pm \iota)}.
\end{equation}

The field associated with Eq.\ (\ref{eq:elementcolor}) is not
truly evanescent because it increases with distance and an
evanescent field should tend to zero. This fact explains the
non-existence of isolated quarks \cite{Halzen84}. In reality even
the gravitational and electric fields are not truly evanescent,
as they tend asymptotically to oscillatory fields of unit
amplitude. This is seen as a feature that ensures compatibility
with the uncertainty principle and fills vacuum with the extreme
richness of the gravitational and electric field remnants from
all the particles in the Universe. It is also a reversal of the
zero point field argument \cite{Haisch01, Nernst16} because mass
and electric charge are the sources of the vacuum field and not
the reverse.
\section{Evanescent fields and worldlines}
The way in which exponential gravitational fields are actually
compatible with Newtonian mechanics has already been explained
\cite{Almeida01:5}, as was the duality between wave description
of movement under gravity and geodesics in 4DO. The approach used
in that paper was less general than is used here because geodesics
were the point of departure and waves were inferred. In Ref.\
\cite{Almeida01:4} the author introduced electromagnetism in the
movement metric of 4DO without resorting to the Euler angles and
this is now seen to be an improper way of facing the problem. In
this paragraph it will be shown that movement under gravitational
or electric fields can easily be derived from the evanescent
fields described above. Magnetic fields will not be treated here
as they have already been shown to result naturally from electric
fields when the source charges are moving \cite{Almeida01:4}.
Movement under color charge fields will only be discussed in
general terms.

When dealing with particles' worldlines it is considered that the
metric is affected by the amplitude of the local field. So, if a
stationary particle with mass $M$ is the field source, using Eq.\
(\ref{eq:elementgrav}), one sets
\begin{equation}
    \label{eq:gravmetric}
    n_{\alpha \beta} = \mathrm{e}^{2 M/r} \delta_{\alpha \beta}.
\end{equation}

For the moment consider a general gravitational field given by
$n_{\alpha \beta}=n^2 \delta_{\alpha \beta}$. Suppose the
particle under the gravitational field's influence is an
electron; the argument is easier to established for an elementary
particle but can be extended to massive bodies as has been
discussed \cite{Almeida01:5}. Until the field equations have been
solved there is no way of knowing what $f(r)$ in Eq.\
(\ref{eq:electron}) looks like but assume that $\Psi$ can be
replaced by a plane wave of amplitude $m$, the electron's mass.
Eq.\ (\ref{eq:wave}) becomes
\begin{equation}
    \label{eq:gravwave3}
    \delta^{\alpha \beta}\partial_\alpha \left(\frac{
    \partial_\beta \Psi}{n^2} \right) = - m^2 \Psi.
\end{equation}
From this equation one derives first of all
\begin{equation}
    \label{eq:gravwave4}
    \delta^{\alpha \beta}\partial_\alpha \left(
    \frac{1}{n^2}\right)\partial_\beta \Psi
    +\frac{\delta^{\alpha
    \beta} \partial_{\alpha \beta}\Psi}{n^2} = - m^2 \Psi.
\end{equation}

The first term of the equation is orthogonal to $\Psi$, in the
sense that $\partial_\beta \Psi$ is something multiplied by
$\mathrm{j}\Psi$. This term precludes the use of a plane wave as
solution to the equation unless, as a limiting case, $n$ is
constant. If $n$ is a slowly varying function, though, one can
neglect the first term and try a solution described by the
function $\Psi = m \exp(\mathrm{j}p_\alpha x^\alpha)$, where
$p_\alpha$ is the wave vector. Eq.\ (\ref{eq:gravwave4}) becomes
\begin{equation}
    \label{eq:gravmoment}
    \frac{1}{m^2 n^2}~ \delta^{\alpha\beta} p_\alpha p_\beta = 1.
\end{equation}

The equation above is the geodesic equation of the space whose
metric is $g_{\alpha \beta} = m^2 n^2 \delta_{\alpha \beta}$ and
can also be written \cite{Almeida01:5}
\begin{equation}
    \label{eq:gravgeod}
    m^2 n^2 \delta_{\alpha \beta}\dot{x}^\alpha \dot{x}^\beta =1,
\end{equation}
where the ''dot'' is used to represent derivatives with respect
to arc length, this being the same as time length in 4DO. The
geodesic and movement Lagrangian is made equal to $1/2$ and
$p_\alpha = m^2 n^2 \delta_{\alpha \beta}\dot{x}^\beta$ is the
conjugate momentum.

From the Lagrangian it is easy to derive the Euler-Lagrange
movement equations:
\begin{equation}
    \label{eq:graveuler1}
    \dot{p}_\alpha = m^2 n \partial_\alpha n \delta_{\mu \nu}
    \dot{x}^\mu \dot{x}^\nu.
\end{equation}
Considering (\ref{eq:gravgeod}), $m^2 n \delta_{\mu \nu}
    \dot{x}^\mu \dot{x}^\nu$ can be replaced by $1/n$; making the replacement
\begin{equation}
    \label{eq:graveuler2}
    \dot{p}_\alpha = \partial_\alpha \left(\log n\right).
\end{equation}
Usually $n$ will not depend on $\tau$, with the consequence that
$p_0 = m^2 n^2 \dot{x}^0$ is a constant of the movement.

Consider now the field due to a particle of mass $M$, given by
Eq.\ (\ref{eq:gravmetric}), insert into Eq.\ (\ref{eq:graveuler2})
and re-write in spherical coordinates to get the 4 differential
equations of orbital motion. Making $\theta=\pi/2$ it is:
\begin{itemize}
    \item $\mathrm{e}^{2 M/r} \dot{\tau} = \gamma$, constant,
    $\rightarrow$
    speed limited by the speed of light;
    \item $\mathrm{e}^{2 M/r} r^2 \dot{\varphi} = J$, constant,
    $\rightarrow$
    perihelium advance in closed orbits;
    \item $\displaystyle{\dot{p}_r = -\frac{M}{r^2}  +
    \frac{m^2 \mathrm{e}^{-2M/r}
    J^2}{r^3}}$, $\rightarrow$
    compatibility with Newton mechanics.
\end{itemize}

The situation is not very different in the case of an electric
field, where one has to consider a field metric of the type
$n_{\alpha \beta} = \delta_{\alpha \beta}$ when the two indices
are not simultaneously zero and $n_{00} = n^2$. In particular,
the field originated by a point charge $Q$ results in $n =
\exp(Q/r)$; inserting into Eq.\ (\ref{eq:wave}) one gets
\begin{equation}
    \label{eq:chargewave2}
   \frac{
    \partial_{0 0} \Psi}{n^2}+ \delta^{i j} \partial_{i j}
    \Psi = - m^2 \Psi,
\end{equation}
with $i,j = 1$ to $3$.

Inserting $\partial_\alpha \Psi = \mathrm{j} p_\alpha \Psi$
\begin{equation}
    \label{eq:chargemoment}
    \frac{\left(p_0\right)^2}{m^2 n^2} + \frac{\delta_{i j} p_i
    p_j}{m^2} = 1,
\end{equation}
which is a geodesic equation of the space and can equivalently be
written
\begin{equation}
    \label{eq:chargegeod}
    m^2 n^2 \left(\dot{x}^0\right)^2 + \delta_{i j}\dot{x}^i
    \dot{x}^j =1.
\end{equation}
Proceeding as above one can set the Lagrangian equal to $1/2$.

If $n$ is a function of the spatial coordinates only, $p_0 = m^2
n^2 \dot{x}^0$ is constant; for the other components one gets
\begin{equation}
    \label{eq:chargecanon}
    \dot{p}_i = m^2 n \partial_i n \left(\dot{x}^0 \right)^2 =
    p_0 \partial_i \left(\log n\right).
\end{equation}
This equation represents movement under an electric field if
$p_0$ equals the electric charge of the moving particle. Notice
that the equation above is not entirely coincident with the
equation used in Ref.\ \cite{Almeida01:4} for a similar
situation, where the approach was different and now considered
incorrect.

When gravitational and electric fields are considered
simultaneously one has to associate to the moving particle a
field with two components: gravitational and electric. Each of two
components verifies Eqs.\ (\ref{eq:gravwave3}) and
(\ref{eq:chargewave2}) respectively. The field to be considered
is then a two-component vector field and the appropriate wave
equation is a vector wave equation. The two component vector
field is an artifact to represent the ''real'' twisted field
through the decomposition into ''longitudinal'' and ''torsional''
components.

Quark interaction needs special attention and will not be
discussed here. Suffice it to say that either a quark and an
anti-quark or three quarks of different color must be associated
in order to cancel most or all the color filed and create a
''stable'' particle. The association does not imply annihilation,
on the contrary, the associated quarks orbit around each other
greatly contributing with this orbital motion to the total mass.
Consideration of color charge implies that the total number of
vector field components is increased from two to four, because
isotropic rotation can be associated with simultaneous influence
on the three charge field components.
\section{Interactions and carrier particles}
The interaction between two particles can be understood as
coupling between the fields in two waveguides through their
respective evanescent fields. This process generates the force of
gravity, electric force and also color force without the need to
appeal to force carriers. One can postulate virtual carriers such
as gravitons, photons and gluons, respectively, but this is a
wave process that does not require such mechanism. For massive
bodies with a large number of internal modes that are allowed to
move in response to mutual interactions the worldlines of their
centers of mass become bent and so do their respective waveguides,
allowing effective mode exchange between them. This phenomenon
happens very dramatically in atom excitation and decay but it
happens also, in a less dramatic form, with the very closely
spaced modes of planet orbits \cite{Almeida01:5}.

A different situation happens with carrier particles that exist on
their own, of which one knows for sure there are photons and
assumes there can also be gluons and gravitons. One must start by
defining carrier particles in some way that is consistent with
observations and with the theory developed above. Carrier
particles are not expected to result from guided modes of the
Higgs field and are expected to have zero momentum along the
$\tau$ direction. Using photons as the only available example to
test with observations, it is clear that their worldlines must
exist in 3-dimensional space \cite{Almeida01:4, Almeida02} and so
their momentum is characterized by $p_0=0$. If this condition is
inserted in Eq.\ (\ref{eq:wave}) one gets exactly the same
equation, with indices varying between 1 and 3.

Just as in the 4-dimensional case, the equation will generate
self-guided modes which will exist completely in 3-dimensional
space. These modes don't have the richness of their 4-dimensional
counterparts and are very similar to ordinary optical fibers. The
field that undergoes self-guidance is the electric field in the
case of photons and the gravitational field in the case of
gravitons. Presumably gluons can be explained similarly if the
analysis is extended to the color field. Without the constraint
of a Universal field, carrier particles can be generated with any
longitudinal frequency, unlike massive particles which are
restricted to the modes supported by the Higgs field.
\section{Conclusions}
A Universal wave equation is shown to provide self-guidance to
the postulated Higgs field in 4DO. 4-dimensional waveguides
result with modes that can explain all the elementary particles
such as are understood today. Although solutions to the
differential equations that result could not be found, the
rationale that is presented supports the argument towards this
being a plausible model.

Waveguiding originates evanescent fields in 4-dimensions such as
they do in 3-dimensional optical fibers; these are shown to be
responsible for gravitational and electric fields and it is
argued that a similar effect can explain color fields. Forces
result from coupling between waveguides of different particles by
intermediate evanescent fields. The equivalence between wave
description and worldlines is explored for the cases of gravity
and electric field, justifying the introduction of a vector field
concept to describe elementary particles.

Carrier particles are then dealt with by setting the $\tau$
component of moment equal to zero. The wave equation becomes
3-dimensional and self-guidance then explains the appearance of
photons and gravitons. Gluons are not discussed, although it is
expected that they can be explained similarly.
%
  \bibliography{aberrations}   
  \bibliographystyle{spiebib}

\end{document}